\documentclass[iop]{emulateapj}

\shorttitle{New Standard} \shortauthors{Contopoulos et al.}

\def\gsim{\mathrel{\raise.5ex\hbox{$>$}\mkern-14mu
             \lower0.6ex\hbox{$\sim$}}}
\def\lsim{\mathrel{\raise.3ex\hbox{$<$}\mkern-14mu
             \lower0.6ex\hbox{$\sim$}}}

\begin{document}

\author{Ioannis Contopoulos\altaffilmark{1,*},
Constantinos Kalapotharakos\altaffilmark{2,3} and Demosthenes
Kazanas\altaffilmark{3}}

\title{A new standard pulsar magnetosphere}

\altaffiltext{1}{Research Center for Astronomy, Academy of Athens,
Athens 11527, Greece}
\altaffiltext{2}{University of Maryland,
College Park (UMDCP/CRESST), College Park, MD 20742, USA}
\altaffiltext{3}{NASA/GSFC, Code 663, Greenbelt, MD 20771, USA}
\altaffiltext{*}{icontop@academyofathens.gr}


\begin{abstract}In view of recent efforts to probe the physical
conditions in the pulsar current sheet, we revisit the standard
solution that describes the main elements of the ideal force-free
pulsar magnetosphere. The simple physical requirement that the
electric current contained in the current layer consists of the
local electric charge moving outward at close to the speed of
light, yields a new solution for the pulsar magnetosphere
everywhere ideal force-free except in the current layer. The main
elements of the new solution are a) the pulsar spindown rate of
the aligned rotator is $23\%$ times larger than that of the
orthogonal vacuum rotator, b) only $60\%$ of the magnetic flux
that crosses the light cylinder opens up to infinity, c) the
electric current closes along the other $40\%$ which gradually
converges to the equator, d) this transfers $40\%$ of the total
pulsar spindown energy flux in the equatorial current sheet which
is then dissipated in the acceleration of particles and in
high-energy electromagnetic radiation, e) there is no separatrix
current layer. Our solution is a minimum free-parameter solution
in that the equatorial current layer is electrostatically
supported against collapse and thus does not require a thermal
particle population. In that respect, it is one more step toward
the development of a new standard solution. We discuss the
implications for intermittent pulsars and long duration gamma-ray
bursts. We conclude that the physical conditions in the equatorial
current layer determine the global structure of the pulsar
magnetosphere.
\end{abstract}



\section{Introduction}

The current theory of pulsars is founded on the assumption that
the magnetosphere of a rotating magnetic neutron star will be
filled with an electron-positron plasma that everywhere satisfies
the force-free and ideal MHD conditions $\rho {\bf E}+{\bf J\times
B}=0$ and ${\bf E\cdot B}=0$ respectively (Goldreich \& Julian
1969). Here, ${\bf E}$ and ${\bf B}$ are the magnetospheric
electric and magnetic fields, and $\rho\equiv \nabla\cdot {\bf E}$
and ${\bf J}$ are the magnetospheric electric charge and current
densities as measured by an inertial (non-rotating) observer.
Based on this assumption, it has been shown that magnetic field
lines cannot remain closed beyond the so-called light cylinder
radius $R_{\rm LC}\equiv c/\Omega$ where $\Omega$ is the neutron
star angular velocity, and that open magnetic field lines carry
the electromagnetic energy responsible for the pulsar spindown.
Contopoulos, Kazanas \& Fendt~(1999; hereafter CKF) first showed
that the Goldreich-Julian requirements cannot be implemented
everywhere in the magnetosphere. When one tries to obtain a
solution that fills all space (and not one with regions of
avoidance as in Lovelace, Turner \& Romanova~2006), he/she finds
that it necessarily involves an equatorial current sheet which
provides closure for the large scale electric current that flows
from the star to infinity and back. In particular, the equatorial
current sheet first discovered by CKF continues all the way to the
stellar surface through two separatrix current sheets that meet at
the so-called Y-point where the region of closed field lines ends
on the light cylinder. This has now become the `standard' picture
of the pulsar magnetosphere.

Unfortunately, the CKF solution (and all subsequent improvements
and generalizations of it in ideal MHD) cannot say anything about
how it is implemented microphysically. In other words, the
solution for ${\bf E}$ and ${\bf B}$ requires a uniquely
determined distribution of $\rho$ and ${\bf J}$, only there is no
provision in the solution for where the required charges come from
and how they are channelled around in the magnetosphere. This
problem is most severe along the equatorial magnetospheric current
sheet in the interior of which ideal MHD fails and dissipation of
electromagnetic energy most probably takes place. Therefore, in
order to understand how the ideal MHD global solution is
implemented in practice, we need to focus on the physical
conditions in the equatorial current sheet where the physical
requirements are most severe.

Indeed, there has been strong recent interest in the pulsar
current sheet microphysics (Coroniti 1990; Lyubarski \& Kirk 2001;
Uzdensky, Cerutti \& Begelman 2011; Uzdensky \& Spitkovsky 2012;
Cerutti, Uzdensky, \& Begelman 2013). All these studies focus on
the current sheet microphysics {\em independently of} the global
structure of the magnetosphere. In a series of 3 papers
(Contopoulos~2007abc), we emphasized the strong coupling between
the global magnetic field topology and reconnection in the
equatorial current sheet. We showed that the non-ideal MHD
condition in the current sheet may dramatically modify the global
electric current distribution and consequently the global
magnetospheric solution too. These results are confirmed in recent
efforts to move away from ideal MHD with the introduction of
magnetospheric dissipation through various ad hoc prescriptions
for a finite plasma conductivity (Kalapotharakos, Kazanas, Harding
\& Contopoulos 2012; Li, Spitkovsky \& Tchekhovskoy 2012b).

In view of the recent efforts to determine the physical conditions
in the pulsar current layer, and motivated by our earlier 2007
work, we believe that there is need to reconsider the main
elements of the ideal force-free pulsar magnetosphere. We realized
that by imposing a simple microphysical requirement in the
equatorial current sheet, we uniquely obtain {\em a new pulsar
magnetosphere solution} which may be used as {\em the new
standard} for calculating pulsar spindown, magnetospheric particle
acceleration, and magnetospheric radiation emission.

\section{The Old Standard Solution}

In the present work we consider only the axisymmetric case. We
will work in a cylindrical system of coordinates $(R,\phi,z)$
centered on the pulsar and aligned with its rotation axis with
$\phi$ an ignorable coordinate. We find it convenient to rescale
our radial coordinate to $x\equiv R/R_{\rm LC}$. The magnetic axis
may be aligned/counter-aligned with the rotation axis, in which
case we will call the pulsar an aligned/counter-aligned rotator.
We also introduce here the function $\Psi=\Psi(x,z)$ as the
magnetic flux contained inside cylindrical radius $x$ at position
$(x,z)$. Let us begin with a short review of the main
microphysical issues of the CKF solution where current sheets are
treated as dissipationless tangential discontinuities:
\begin{enumerate}
\item The distribution of poloidal electric current
\[
I_{\rm CKF}(\Psi)\approx \left\{
\begin{array}{ll}
\frac{\Omega\Psi}{4\pi}\left(2-\frac{\Psi}{\Psi_{\rm
open}}\right)&
\mbox{if $\Psi \leq \Psi_{\rm open}$}\\
0 & \mbox{if $\Psi> \Psi_{\rm open}$}
\end{array}
\right. \] along open magnetic field lines is obtained as an
eigenfunction of the problem by requiring that the solution
crosses the light cylinder smoothly. Here, $\Psi_{\rm open}=1.23
\Psi_{\rm dipole}$, where $\Psi_{\rm dipole}\equiv \pi r_*^3
B_*/R_{\rm LC}$, $B_*$ is the polar value of the neutron star
magnetic field, and $r_*$ is its radius. This contains an
equatorial current sheet beyond the light cylinder ($x>1$) with a
surface charge density
\begin{eqnarray}
\sigma(x,0) & = & 2E_{z}(x,0^+)\nonumber\\
& = & 2xB_R(x,0^+)\ .
\end{eqnarray}
$\sigma$ is positive/negative, therefore, the current sheet
consists mostly of outflowing positrons/electrons. Their origin
may be related to pair formation at the Y-point. Note that,
\begin{equation}
\sigma(1^+,0)=0
\end{equation}
right outside the magnetospheric Y-point on the light cylinder,
since $B_R(1^+,0)=0$ there. The fact that the equatorial current
sheet does not only carry an electric current but is also charged
has not been adequately emphasized in previous studies. As we will
see below, this has some very important consequences for the
global structure of the pulsar magnetosphere. \item The equatorial
current sheet closes onto the star through two current sheets
along the separatrix between open and closed field lines. The
separatrix current sheets are charged with a surface charge
density
\begin{eqnarray}
\sigma(x,z) & \equiv &
E(x,z^+)-E(x,z^-)\nonumber\\
& = & -x B_p(x,z^+)\cdot\nonumber
\end{eqnarray}
\begin{equation}
\left\{\left[1+\frac{[2I(\Psi_{\rm
open})/(Rc)]^2}{x^2[1-x^2]B_p^2(x,z^+)}\right]^{1/2}-1\right\}\ .
\end{equation}
$\sigma$ is negative/positive, therefore, the separatrix current
sheets consist mostly of inflowing
electrons/positrons\footnote{For the history of this discovery, we
must acknowledge that although CKF did indeed discover the
magnetospheric current sheet associated with the poloidal
magnetospheric current, it was the referee of their paper, Pr. Leo
Mestel, who pointed out the presence of tangential discontinuities
at the two separatrices.}. Their origin too may be related to pair
formation at the Y-point (Contopoulos 2009). Here, $B_p$ is the
poloidal magnetic field. \item Open field lines that cross the
zero space-charge (null) surface contain a distributed
outflowing/inflowing electric current consisting mostly of
inflowing electrons/positrons inside the null surface and
outflowing positrons/electrons outside. Their origin may be
related to pair cascades in the so-called outer gaps. \item A
small amount of open field lines cross the null surface twice and
contain a distributed inflowing/outflowing electric current. The
origin of its associated charge carriers is obscure. \item Most
open field lines around the pole do not cross the null surface.
Those contain a distributed inflowing/outflowing electric current
consisting mostly of outflowing electrons/positrons. Their origin
may be related to pair cascades on the pulsar polar cap or in the
outer gaps.
\end{enumerate}

The most important part of the magnetosphere is the equatorial
current sheet which guarantees the global poloidal electric
current closure. The current sheet experiences a pinching force
due to its reversing surface magnetic field components
$B_\phi(x,0^+)=-B_\phi(x,0^-)$ and $B_R(x,0^-)=-B_R(x,0^-)$. The
new element that we want to emphasize here is that the current
sheet is also charged because the electric field component
perpendicular to it also reverses across it, i.e.
$E_z(x,0^+)=-E_z(x,0^-)$. Therefore, the vertical electrostatic
self-repulsion of the current sheet surface charge density
$\sigma$ contributes to its vertical support against the magnetic
pinching force. Note that, in contrast to the solar corona, there
is no `guide field' in the pulsar reversing equatorial current
sheet as some recent investigations assume (Cerruti, Uzdensky \&
Begelman 2013).

\section{The Equatorial Current Sheet}

Let us now study the small scale structure of the current sheet
and assume that it has a thickness $2h\ll R_{\rm LC}$. In the
current sheet interior defined as $-h\leq z\leq h$, in general
\begin{eqnarray}
B_R(x,z) & = & \frac{z}{h}B_R(x,h)\\
B_z(x,z) & = & B_z(x,h)\\
B_\phi(x,z) & = & \frac{z}{h}B_\phi(x,h)\\
E_R(x,z) & = & E_R(x,h)=-xB_z(x,h)\label{ER}\\
E_z(x,z) & = & \frac{z}{h}E_z(x,h)=xB_R(x,z)\\
E_\phi(x,z) & = & 0
\end{eqnarray}
Note that $E_R(x,0)=B_z(x,0)=0$ in CKF. In general,
$E_R(x,0)=xB_z(x,0)> B(x,0)$. We then have,
\begin{eqnarray}
{\bf J}_{\rm cs} & = & -\frac{B_\phi(x,h)c}{h}\hat{\rm
R}+\frac{B_R(x,h)c}{h}\hat{\phi}\ ,\\
\sigma_{\rm cs} & = & 2E_z(x,h)=2xB_R(x,h)\ ,\\
\rho_{\rm cs} & = & \frac{E_z(x,h)}{h}=\frac{xB_R(x,h)}{h}\ .
\end{eqnarray}
The vertical force experienced by the current sheet plasma is
equal to
\begin{eqnarray}
F_z(x,z)_{\rm cs} & \equiv & \rho E_z + \frac{J_R B_\phi - J_\phi
B_R}{c} \nonumber
\\ & = &
\frac{[(x^2-1)B_R^2(x,h)-B_\phi^2(x,h)]z}{h^2}
\label{Fz}
\end{eqnarray}
In the old standard pulsar magnetosphere solutions, $zF_z$ is
negative, i.e. the inward magnetic pinching force dominates. Most
importantly though, at the points where the equatorial current
sheet meets the light cylinder,
\begin{eqnarray}
F_z(1,z) & = & -\frac{B_\phi^2(1,h)z}{h^2}\ .\
\end{eqnarray}
In standard solutions, $B_\phi(1,h)=2I(\Psi_{\rm open})/(Rc)\neq
0$, and therefore, $F_z(1,z)\neq 0$. An equivalent way to say that
the magnetic pinching force in the current sheet dominates is to
realize that the electric current in the current sheet in the old
standard solution is {\em spacelike}, i.e. that
\begin{eqnarray}
J_{\rm cs} & = & \frac{(B_\phi^2(x,h)+B_R^2(x,h))^{1/2}c}{h} \nonumber\\
& > & \frac{E_z(x,h)c}{h}= \rho_{\rm cs} c\ . \label{spacelike}
\end{eqnarray}
This picture requires extra charge carriers of the opposite sign
(electrons/positrons) in excess of the ones needed to account for
$\rho_{\rm cs}$. It also requires a `hot' population of particles
to support the current sheet against its own magnetic pinching.

A very simple and natural simplification to our problem is to
assume that the electric current in the current sheet is {\em
everywhere null}, or at least only {\em weakly spacelike}. In that
case, eq.~(\ref{Fz}) yields $F_z\approx 0$. In other words, a null
equatorial current sheet is electrostatically supported by its own
charge and requires no additional `hot' population of particles.
The above requirement amounts to a new equatorial boundary
condition given implicitly through
\begin{equation}
\frac{2\Omega I(\Psi(x,h))}{c^2}=x(x^2-1)^{1/2}B_R(x,h)
\end{equation}
or equivalently
\begin{equation}
\frac{\Omega I(\Psi(x,h))}{\pi c^2}=(x^2-1)^{1/2}\left.
\frac{\partial\Psi(x,z)}{\partial z}\right|_{z=h} \label{newBC}
\end{equation}
instead of the `standard' CKF equatorial condition $\Psi(x,h)=
\Psi(x,0)= \Psi(1,0)\equiv \Psi_{\rm open}$. Note that
$2I(\Psi(x,h))/(Rc)=xB_\phi(x,h)$.

\section{The New Standard Solution}

As described in CKF and Timokhin~(2006), we solve the pulsar
equation numerically by iteratively relaxing both the distribution
of $\Psi(x,z)$ everywhere in the half space $z>0$ (given the
boundary conditions along the axis, infinity, and the equator),
and the distribution of magnetospheric electric current $I(\Psi)$
that allows for a smooth and continuous crossing of the light
cylinder. The main difference with CKF is the new equatorial
boundary condition (eq.~\ref{newBC}) which is implemented during
the iteration to evolve the equatorial distribution $\Psi(x,0)$
for the next iteration step.


\begin{figure}
\centering
\includegraphics[trim=0cm 0cm 0cm 0cm, clip=true,
width=10cm, angle=0]{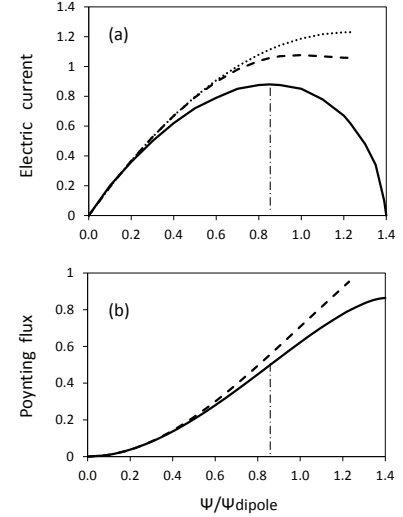} \caption{(a) Distribution of
electric current  $4\pi I_{\rm New}/c\Psi_{\rm dipole}$ as a
function of $\Psi/\Psi_{\rm dipole}$ along `open' field lines.
$\Psi_{\rm open}=1.4\Psi_{\rm dipole}$. $I_{\rm New}(\Psi_{\rm
open})=0$. Solid/dashed/dotted lines: New/CKF/monopole solution
respectively. Dashed-dotted line: marks the value of $\Psi_{\rm
max\ I}=0.85\Psi_{\rm dipole}=0.60\Psi_{\rm open}$ beyond which
the return magnetospheric current flows; (b) Distribution of
Poynting flux normalized to $\dot{E}_{\rm CKF}$. Solid/dashed
lines: New/CKF solution respectively. Dashed-dotted line marks the
value of $\Psi_{\rm max\ I}$ beyond which the magnetospheric
Poynting flux gradually enters the current sheet.}
\end{figure}

\begin{figure}
\centering \centering
\includegraphics[trim=6.5cm 0cm 6.5cm 0cm, clip=true,
width=7.38cm, angle=0]{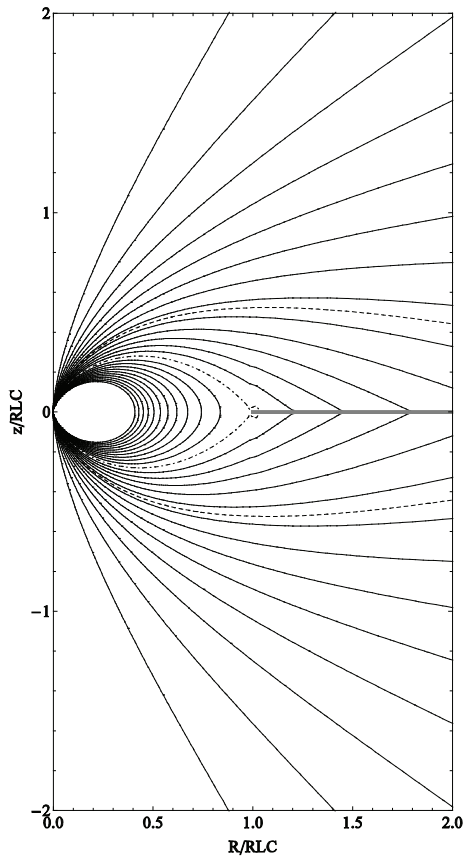} \caption{The new structure of
the pulsar magnetosphere with a null equatorial current sheet.
Shown are magnetic flux ($\Psi$) surfaces plotted in intervals of
$0.1 \Psi_{\rm dipole}$. Dashed line: $\Psi_{\rm max\
I}=0.85\Psi_{\rm dipole}$ (corresponds to the boundary between
outgoing and ingoing magnetospheric current, and asymptotically
approaches the null line -i.e. the boundary between positive and
negative space charge- and the equator as $R\rightarrow\infty$).
Dashed-dotted line: $\Psi_{\rm open}=1.4\Psi_{\rm dipole}$. Thick
gray line: equatorial current sheet.}
\end{figure}

\begin{figure}
\centering \centering
\includegraphics[trim=6.5cm 0cm 6.5cm 0cm, clip=true,
width=7.5cm, angle=0]{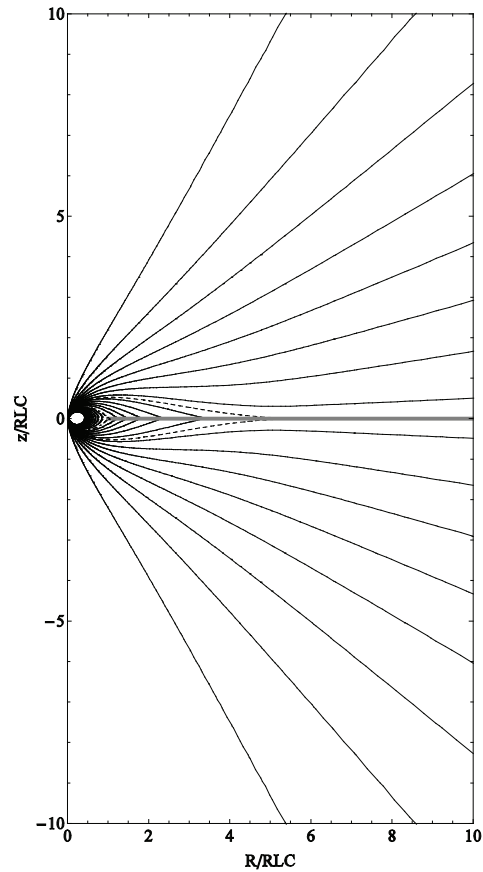} \caption{Same as figure~2 but
on a larger scale.}
\end{figure}

Our results are shown in figures~1-3. In figure~1a we show the
distribution of poloidal electric current that we obtained along
field lines that cross the light cylinder (we will still call them
`open'). This can be fitted  with the analytic expression
\[
I_{\rm New}(\Psi)\approx \]
\[ \left\{
\begin{array}{ll}
\frac{1.07\Omega\Psi}{4\pi}\left(2-\frac{\Psi}{\Psi_{\rm
open}}\right)\left(1-\frac{\Psi}{\Psi_{\rm open}}\right)^{0.4} &
\mbox{if $\Psi<\Psi_{\rm open}$}\\
0 & \mbox{if $\Psi\geq \Psi_{\rm open}$}
\end{array}
\right. \] Here, $\Psi_{\rm open}=1.4 \Psi_{\rm dipole}$. For
comparison, we also show the CKF current distribution and its
corresponding monopole distribution. In figure~1b we show the
distribution of Poynting flux in the pulsar magnetosphere,
starting from the axis ($\Psi=0$) up to the last open field line
($\Psi=\Psi_{\rm open}$). For comparison, we also show the CKF
Poynting flux distribution.

The main elements of the new solution are a) the pulsar spindown
rate is equal to
\begin{equation}
\dot{E}=0.82\frac{B_*^2 \Omega^4 r_*^6}{4c^3} \label{Edot}
\end{equation}
i.e. $82\%$ of the CKF spindown, or $23\%$ higher than that of the
orthogonal vacuum rotator, b) $60\%$ of the magnetic flux that
crosses the light cylinder, $\Psi_{\rm max\ I}$, carries the one
part of the magnetospheric current that consists of
electrons/positrons that outflow to infinity, c) the electric
current closes along the other $40\%$ that consists of
positrons/electrons that gradually converge to the equator, d)
this transfers $40\%$ of the total pulsar spindown energy flux in
the equatorial current sheet which is then dissipated in the
acceleration of particles and in high-energy electromagnetic
radiation within a few times the light cylinder
radius\footnote{$90\%$ of the dissipated energy is dissipated
within 6 light cylinder radii.}, e) there is no separatrix current
sheet, f) pressure balance across the equatorial current layer is
guaranteed by the electrostatic repulsion of the charge contained
in it. Our solution is a minimum free-parameter solution in that
the equatorial current sheet does not require a thermal particle
population for vertical support against collapse.

The most important element of that solution is e), namely the fact
that
\begin{equation}
I_{\rm New}(\Psi_{\rm open})=0 \label{Inew}
\end{equation}
(eq.~\ref{newBC} for $x=1$), and $B_\phi(1,h)=0$. In other words,
the electric current closes within the field lines that cross the
light cylinder (we will still call them `open') and thus there is
no need for a current sheet along the separatrix between open and
closed field lines. The equatorial current sheet is established to
provide closure for the global pulsar electric circuit. As we
said, only magnetic field lines that carry the return current
consisting of outflowing positrons/electrons enter the current
sheet, and then outflow along the current sheet to infinity. The
main magnetospheric current consists of electrons/positrons that
outflow along the other $60\%$ of magnetic field lines that
originate closer to the polar cap axis. It is interesting that
eq.~(\ref{Inew}) naturally resolves a complication in the CKF
solution, namely that $\left. B_p(x,0)\right|_{x\rightarrow 1^-}=
\left. 2I_{\rm CKF}(\Psi_{\rm open})\Omega/c^2
[x^2-1]^{1/2}\right|_{x\rightarrow 1^-}\rightarrow \infty$
(Uzdensky~2003; Contopoulos 2009).

Note the qualitative similarity between the solution shown in
figures~2 \& 3 and the solution obtained in figure~5 of
Contopoulos~(2007c) which was based on analogous qualitative
global arguments. In that solution, though, there was no provision
for the ultimate opening up of field lines that carry the main
magnetospheric current, and therefore the associated spindown
rates obtained in that work are different.

\section{Discussion}

We have presented here a solution that we hope will become the new
standard for calculating pulsar spindown, magnetospheric particle
acceleration, and magnetospheric radiation emission. It is the
first global solution that self-consistently takes care of the
equatorial current sheet microphysical conditions in the limit of
a null current. In that limit, the current sheet is self-supported
electrostatically, and in that respect it is a `minimum parameter'
solution, since it does not require an extra population of hot
positrons/electrons. We must acknowledge that, the electromagnetic
(Poynting) energy that is deposited in the current sheet is
dissipated in the acceleration of positrons/electrons by the
$E_R=xB_z>B\neq 0$ electric field component in the current sheet
(eq.~\ref{ER}). Obviously, some fraction of those particles is
expected to form a hot population which will partially contribute
to the electrostatic support of the current sheet. In that
respect, our present solution must be considered as one more step
toward the development of a new standard solution. It is
interesting though that, since particle acceleration proceeds {\em
beyond} the light cylinder, any hot population will also form {\em
beyond} the light cylinder. The absence of a hot population at the
Y-point justifies one of the main results of the present work,
namely the elimination of the separatrix current sheet.

The electrons and positrons that populate the pulsar magnetosphere
may naturally be produced through pair production at the crossings
of the zero space charge (null) surface, the so called outer gaps.
The electrons/positrons that move inward continue to flow through
the polar cap and subsequently outflow in the main part of the
polar cap, whereas the positrons/electrons that move outward reach
the equatorial current sheet beyond the light cylinder (these are
the ones that absorb the electromagnetic energy dissipated in the
current sheet). Our minimum parameter solution is physically
appealing in that it guarantees that pair production is not needed
anywhere else in the magnetosphere.

Our solution addresses a fundamental issue that other ideal
force-free solutions cannot address, namely what is the fraction
of the released electromagnetic energy that is deposited in the
current sheet and is dissipated in the acceleration of current
sheet positrons/electrons. Our numerical integration yields
$40\%$\footnote{That value will change when we incorporate partial
thermal support in the current sheet beyond the light cylinder.},
and it is expected that some significant fraction of it will be
seen as high energy pulsar radiation. We have thus found the main
region of magnetospheric dissipation: {\em the equatorial current
sheet within a few light cylinder radii beyond the light
cylinder}.

Our results may also find application in the phenomenon of
intermittent pulsars who are observed to spin down faster during
their ON state than during their OFF state. If we associate the ON
state with our present solution, the spin down rate of the ON
state is only $82\%$ of that of the CKF solution. We can naively
generalize eq.~(\ref{Edot}) in 3D in analogy to Spitkovsky~(2006)
as
\begin{equation}
\dot{E}_{\rm ON}(\alpha)\sim 0.82\frac{B_*^2 \Omega^4
r_*^6}{4c^3}(1+\sin^2\alpha)\ , \label{EdotON}
\end{equation}
where $\alpha$ is the inclination angle between the magnetic pole
and the rotation axis. If we also associate the OFF state with the
Li, Spitkovsky \& Tchekhovskoy~(2012a) solution with nearly vacuum
conditions on the open field lines and nearly ideal force-free
conditions on the closed field lines (where plasma remains trapped
in the absence of pair production),
we obtain a ratio of respective spin down rates $\dot{E}_{\rm
ON}/\dot{E}_{\rm OFF}~ 1 - 2.5$ for a likely range of pulsar
inclination angles of $90^\circ - 30^\circ$. This may naturally
account for observed values between 1.5 and 2.5.


Our solution may also find application in the study of long
duration gamma-ray bursts (GRB) where the central energy source is
believed to be a newly formed maximally rotating stellar mass
black hole which loses energy through electrodynamic processes
very similar to the ones that take place in the pulsar
magnetosphere (Blandford \& Znajek~1977; Komissarov \&
Barkov~2009; Contopoulos, Nathanail \& Pugliese~2013). In
particular, a maximally rotating $10~M_\odot$ black hole can lose
up to about $10^{55}$~ergs of rotational kinetic energy in the
form of Poynting flux (Contopoulos, Nathanail \& Pugliese~2013).
However, it is still not clear what fraction of that spin down
energy will be radiated in the form of high energy (X-ray,
$\gamma$-ray) radiation. Our new standard solution yields a
fraction around $40\%$ of the spindown luminosity. If that result
is confirmed, it will allow us to use long duration GRBs as {\em
standard candles in Cosmology}.

We conclude that indeed, the physical conditions in the equatorial
current layer significantly modify the global structure of the
pulsar magnetosphere.

\acknowledgements{

This work was supported by the General Secretariat for Research
and Technology of Greece and the European Social Fund in the
framework of Action `Excellence'.}




{}

\end{document}